\RequirePackage{ifpdf}
\ifpdf 
\documentclass[pdftex]{sigma}
\else
\documentclass{sigma}
\fi

\numberwithin{equation}{section}
\newtheorem{thm}{Theorem}[section]

\theoremstyle{definition}

\theoremstyle{remark}

\newcommand{\commentout}[1]{}

\begin{document}

\allowdisplaybreaks

\renewcommand{\PaperNumber}{045}

\FirstPageHeading

\ShortArticleName{Lattice Structure of Connection Preserving Deformations}

\ArticleName{The Lattice Structure of Connection Preserving\\ Deformations for $\boldsymbol{q}$-Painlev\'e Equations I}

\Author{Christopher M. ORMEROD}

\AuthorNameForHeading{C.M. Ormerod}

\Address{La Trobe University, Department of Mathematics and Statistics,
Bundoora VIC 3086, Australia}
\Email{\href{mailto:C.Ormerod@latrobe.edu.au}{C.Ormerod@latrobe.edu.au}}

\ArticleDates{Received November 26, 2010, in f\/inal form May 03, 2011;  Published online May 07, 2011}

\Abstract{We wish to explore a link between the Lax integrability of the $q$-Painlev\'e equations and the symmetries of the $q$-Painlev\'e equations. We shall demonstrate that the connection preserving deformations that give rise to the $q$-Painlev\'e equations may be thought of as elements of the groups of Schlesinger transformations of their associated linear problems. These groups admit a very natural lattice structure. Each Schlesinger transformation induces a B\"acklund transformation of the $q$-Painlev\'e equation. Each translational B\"acklund transformation may be lifted to the level of the associated linear problem, ef\/fectively showing that each translational B\"acklund transformation admits a Lax pair. We will demonstrate this framework for the $q$-Painlev\'e equations up to and including $q$-$\mathrm{P}_{\mathrm{VI}}$.}

\Keywords{$q$-Painlev\'e; Lax pairs; $q$-Schlesinger transformations; connection; isomonodromy}

\Classification{34M55; 39A13}

\section{Introduction and outline}

The discrete Painlev\'e equations are second order non-linear non-autonomous dif\/ference equations admitting the the Painlev\'e equations as continuum limits \cite{Gramani:DiscretePs}. They are considered integrable by the integrability criterion of singularity conf\/inement \cite{Gramani:DiscretePs}, solvability via associated li\-near problems \cite{Gramani:Isomonodromic} and have zero algebraic entropy \cite{Viallet}. There are three classes of discrete Painlev\'e equations: the additive dif\/ference equations, $q$-dif\/ference equations and elliptic dif\/ference equations. This classif\/ication is in accordance with the way in which the non-autonomous variable evolves with each application of the iterative scheme. The aim of this article is to derive a range of symmetries of the considered $q$-dif\/ference Painlev\'e equations using the framework of their associated linear problems.

The discovery of the solvability of a $q$-Painlev\'e equation via an associated linear problem was established by Papageorgiou et al.~\cite{Gramani:Isomonodromic}. It was found that a $q$-Painlev\'e equation can be equivalent to the compatibility condition between two systems of linear $q$-dif\/ference equations, given by
\begin{subequations}
\begin{gather}
\label{qx}Y(qx,t)  = A(x,t)Y(x,t),\\
\label{qt}Y(x,qt)  = B(x,t)Y(x,t),
\end{gather}
\end{subequations}
where $A(x,t)$ and $B(x,t)$ are rational in $x$, but not necessarily $t$, which is the independent variable in the $q$-Painlev\'e equation. A dif\/ference analogue of the theory of monodromy preserving deformations \cite{Jimbo:Monodromy1, Jimbo:Monodromy2, Jimbo:Monodromy3} was proposed by Jimbo and Sakai \cite{Sakai:qP6}, who introduced the concept of connection preserving deformations for $q$-Painlev\'e equations. Given a system of linear $q$-dif\/ference equations, of the form \eqref{qx}, conditions on the existence of the formal solutions and connection matrix associated with \eqref{qx} has been the subject of Birkhof\/f theory \cite{Birkhoffallied, Carmichael, Adams}. The preservation of this connection matrix leads naturally to the auxiliary linear problem in $t$, given by \eqref{qt}. This theory also led to the discovery of the $q$-analogue the sixth Painlev\'e equation in an analogous manner to the classical result of Fuchs \cite{Fuchs1, Fuchs2}.

There was one unresolved issue in the explanation of the emergence of discrete Painlev\'e type systems, and this was related to the emergence, and def\/inition, of $B(x,t)$. Unlike the theory of monodromy preserving deformations and the surfaces of initial conditions where there exists just one canonical Hamiltonian f\/low \cite{Takano:Symplectic, Jimbo:Monodromy1}, there are many ways to preserve the connection matrix. According to the existing framework, one may determine $B(x,t)$ from a known change in the discrete analogue of monodromy data, i.e., the characteristic data. However, we would argue that this known change is not canonical, or unique. That is to say that by considering all possible changes in the characteristic data, one formulates a system of Schlesinger transformations inducing a system of B\"acklund transformations of the Painlev\'e equation. A particular case of this theory was studied by the author in connection with the $q$-dif\/ference equation satisf\/ied by the Big $q$-Laguerre polynomials \cite{studyqPV}.

This concept is not entirely new, one only needs to examine Jimbo and Miwa's second paper~\cite{Jimbo:Monodromy2} to see a set of discrete transformations, known as Schlesinger transformations, specif\/ied for the Painlev\'e equations. These Schlesinger transformations can be thought of as a discrete group of monodromy preserving evolutions, in which various elements of the monodromy data are shifted by integer amounts. The compatibility between the discrete evolution and the continuous f\/low induces a B\"acklund transformation of the Painlev\'e equation, which appear in the associated linear problem, and in some cases, induce the evolution of some discrete Painlev\'e equations of additive type~\cite{Joshi:dP2}.

We will show the same type of transformations for systems of $q$-dif\/ference equations describe similar transformations. We consider systems of transformations in which \eqref{qx} is quadratic. There are only a f\/inite number of cases in which $A(x)$ is quadratic, and these cases cover the $q$-Painlev\'e equations up to and including the $q$-analogue of the sixth Painlev\'e equation \cite{Sakai:qP6}. We will consider the set of transformations of the associated linear problems for the exceptional $q$-Painlev\'e equations in a separate article as we have preliminary results, including a Lax pair for one of these equations that seems distinct from those that have appeared in~\cite{Yamada:LaxqEs} and~\cite{SakaiE6}.

These transformations may be derived in an analogous manner to the dif\/ferential case. If one knows formal expansions of the solutions of \eqref{qx} at $x=0$ and $x=\infty$, then one may formulate the Schlesinger transformations directly in terms of the expansions. Formal expansions for regular $q$-dif\/ference equations are well established in the integrable community \cite{Birkhoffallied, Sakai:qP6}, however, many associated linear problems for $q$-Painlev\'e equations fail to be regular at $0$ or $\infty$. Hence, it is another goal of this article to apply the more general expansions provided by Adams \cite{Adams} and Guenther \cite{BirkhoddAdamsSum} to derive the required Schlesinger equations. These expansions seem to be not as well established, however, all known examples of connection preserving deformations seem to f\/it nicely a framework that encompasses the regular and irregular cases \cite{Sakai:qP6, Murata2009, Gramani:Isomonodromic}. In previous studies, such as those by by Sakai \cite{Sakai:Garnier, Sakai:qP6}, $B(x,t)$ from \eqref{qt} and the Painlev\'e equations seem to have been derived from an overdetermined set of compatibility conditions. In an analogous manner to the continuous case, the formal solutions obtained may or may not converge in any given region of the complex plane~\cite{Trjitzinsky}.

The most important distinction we wish to make here is the absence of a variable $t$. We remark that this should be seen as a consistent trend in the studies of discrete Painlev\'e equations, whereby the framework of Sakai \cite{Sakai:Rational} and the work of Noumi et al.~\cite{Noumi} show us that we should consider the B\"acklund transformations and discrete Painlev\'e equations on the same footing.

The outline of this article is as follows: In Section~\ref{section1} we introduce some special functions associated with $q$-linear systems. In Section~\ref{section2} we shall outline the main results of Birkhof\/f theory as presented by Adams and Guenther~\cite{Adams, BirkhoddAdamsSum} and subsequently, specify the natural lattice structure of the connection preserving deformations. In Section~\ref{section3} we will explore the quadratic matrix polynomial case in depth. Even at this level, we uncover a series of Lax-pairs.

\section[$q$-special functions]{$\boldsymbol{q}$-special functions}\label{section1}

A general theory of special $q$-dif\/ference equations, with particular interest in basic hypergeometric series, may be found in the book by Gasper and Rahman \cite{GasperRahman}. The work of Sauloy and Ramis provide some insight as to how these $q$-special functions both relate to the special solutions of the linear problems and the connection matrix \cite{Sauloy, Ramis}, as does the work of van der Put and Singer \cite{vanderPut, vanderPutSinger}. For the following theory, it is convenient to f\/ix a $q \in \mathbb{C}$ such that $|q| < 1$, in order to have given functions def\/ine analytic functions around $0$ or $\infty$.

We start by def\/ining the $q$-Pochhammer symbol \cite{GasperRahman}, given by
\[
\left( a,q \right)_k = \left\{\begin{array}{ll}
\prod\limits_{n = 1}^{k} \left(1-aq^{n-1}\right) & \mbox{for $k \in \mathbb{N}$},\vspace{1mm}\\
\prod\limits_{n = 1}^{\infty} \left(1-aq^{n-1}\right) & \mbox{for $k = \infty$},\vspace{1mm}\\
1 & \mbox{for $k=0$}. \end{array} \right.
\]
We note that $(a,q)_\infty$ satisf\/ies
\begin{gather*}
(q x;q)_\infty = \dfrac{1}{1-x} (x;q)_\infty,
\end{gather*}
or equivalently,
\begin{gather*}
\left(\dfrac{x}{q};q\right)_\infty = \left( 1- \dfrac{x}{q} \right) (x;q)_\infty.
\end{gather*}

We def\/ine a fundamental building block; the Jacobi theta function \cite{GasperRahman}, given by the bi-inf\/inite expansion
\[
\theta_q(x) = \sum_{n\in\mathbb{Z}} q^{{n \choose 2}} x^n,
\]
which satisf\/ies the $q$-dif\/ference equation
\[
x\theta_q(qx) = \theta_q(x).
\]
The second function we require to describe the asymptotics of the solutions of \eqref{qx} is the $q$-character, given by
\[
e_{q,c}(x)  = \dfrac{\theta_q(x)\theta_q\left(c\right)}{\theta_q \left(xc\right)},
\]
which satisf\/ies
\begin{gather*}
e_{q,c}(qx)  = c e_{q,c}(x),\qquad
e_{q,qc}(x)  = x e_{q,c}(x).
\end{gather*}
Using the above functions, we are able to solve any one dimensional problem.

\section{Connection preserving deformations}\label{section2}

We will take a deeper look into systems of linear $q$-dif\/ference equations of the form
\[
y(q^nx) + a_{n-1}(x)y(q^{n-1}x) + \dots + a_1(x)y(qx) + a_0(x)y(x) = 0 ,
\]
where the $a_i(x)$ are rational functions. One may easily express a system of this form as a matrix equation of the form
\begin{gather*}
Y(qx) = A(x)Y(x),
\end{gather*}
where $A(x)$ is some rational matrix. We quote the theorem of Adams \cite{Adams} suitably translated for matrix equations by Birkhof\/f \cite{BirkhoddAdamsSum} in the revised language of Sauloy et al.~\cite{Sauloy, vanderPut}.

\begin{thm}
Under general conditions, the system possesses formal solutions given by
\begin{gather}\label{fundsols}
Y_0(x)  =  \hat{Y}_0(x)D_0(x), \qquad
Y_\infty(x)  =  \hat{Y}_\infty(x)D_\infty(x),
\end{gather}
where $\hat{Y}_0(x)$ and $\hat{Y}_{\infty}(x)$ are series expansions in $x$ around $x= 0$ and $x=\infty$ respectively and
\begin{gather*}
D_0(x)  = \mathrm{diag} \left(\dfrac{e_{q,\lambda_i}(x)}{\theta_q\left( x\right)^{\nu_i}} \right),\qquad
D_\infty(x)  = \mathrm{diag} \left(\dfrac{e_{q,\kappa_i}(x)}{\theta_q\left(x\right)^{\nu_i}} \right).
\end{gather*}
\end{thm}

Given the existence and convergence of these solutions, we may form the connection matrix, similar to that of dif\/ference equations~\cite{Birkhoff}, given by
\begin{gather*}
P(x) = Y_{\infty}(x) Y_0(x)^{-1}.
\end{gather*}
In a similar manner to monodromy preserving deformations, we identify a discrete set of cha\-racteristic data, namely
\[
M = \left\{\begin{array}{c c c c c c}
\kappa_1 & \ldots & \kappa_n & & &  \\
\lambda_1 & \ldots & \lambda_n & a_1 & \ldots & a_m \end{array}\right\}.
\]
Def\/ining this characteristic data may not uniquely def\/ine $A(x)$ in general. In the problems we consider, which are also those problems considered previously in the work of Sakai~\cite{Sakai:qP6}, Murata~\cite{Murata2009} and Yamada~\cite{Yamada:LaxqEs}, the characteristic data def\/ines a three dimensional linear algebraic group as a system of two f\/irst order linear $q$-dif\/ference equations, but is two dimensional as one second order $q$-dif\/ference equation. The gauge freedom disappears when one extracts the second order $q$-dif\/ference equation from the system of two f\/irst order $q$-dif\/ference equations. Let us suppose the linear algebraic group is of dimension $d$, then let us introduce variables $y_1$, $y_2$, $\ldots$, $y_d$, which parameterize the linear algebraic group. We may now set a co-ordinate system for this linear algebraic group, hence, we write
\begin{gather}\label{defining}
M_A =  \left\{\begin{array}{c c c c c c}
\kappa_1 & \ldots & \kappa_n & & &  \\
\lambda_1 & \ldots & \lambda_n & a_1 & \ldots & a_m \end{array}: y_1, \ldots, y_d \right\},
\end{gather}
as def\/ining a matrix~$A$.

By considering the determinant of the left hand side of~\eqref{qx} and the fundamental solutions specif\/ied by~\eqref{fundsols}, one is able to show
\begin{gather}\label{constraint}
\prod_{i=1}^{n} \kappa_i \prod_{j=1}^{m}(-a_j) = \prod_{i=1}^{n} \lambda_i,
\end{gather}
which forms a constraint on the characteristic data.

We now explore connection preserving deformations \cite{Sakai:qP6}. Let $R(x)$ be a rational matrix in~$x$, then we apply a transformation $Y \to \tilde{Y}$ by the matrix equation
\begin{gather}\label{tildeev}
\tilde{Y}(x) = R(x)Y(x).
\end{gather}
The matrix $\tilde{Y}$ satisf\/ies a matrix equation of the form \eqref{qx}, given by
\begin{gather}\label{assoctilde}
\tilde{Y}(qx) = \left[ R(qx)A(x)R(x)^{-1} \right] \tilde{Y}(x) = \tilde{A}(x)\tilde{Y}(x).
\end{gather}
We note that if $R(x)$ is rational and invertible, then
\begin{gather*}
\tilde{P}(x) =  (\tilde{Y}_\infty(x))^{-1} \tilde{Y}_0(x) = (R(x)Y_\infty(x))^{-1} R(x)Y_0(x) \\
\phantom{\tilde{P}(x)}{} =  (Y_\infty(x))^{-1}R(x)^{-1} R(x)Y_0(x) = Y_\infty(x) Y_0(x),
\end{gather*}
hence, the system def\/ined by
\[
\tilde{Y}(qx) = \tilde{A}(x)\tilde{Y}(x),
\]
is a system of linear $q$-dif\/ference equations that possesses the same connection matrix. However, the left action of $R(x)$ may have the following ef\/fects: The transformation may
\begin{itemize}\itemsep=0pt
\item change the asymptotic behavior of the fundamental solutions at $x = \infty$ by letting $\kappa_i \to q^{n}\kappa_i$;
\item change the asymptotic behavior of the fundamental solutions at $x = 0$ by letting $\lambda_i \to q^{n}\lambda_i$;
\item change the position of a root of the determinant by letting $a_i \to q^na_i$.
\end{itemize}
We use the same co-ordinate system for $\tilde{A}(x)$ as we did for $A(x)$ via \eqref{defining}, giving
\[
M_{\tilde{A}} = \left\{\begin{array}{c c c c c c}
\tilde{\kappa}_1 & \ldots & \tilde{\kappa}_n & & &  \\
\tilde{\lambda}_1 & \ldots & \tilde{\lambda}_n & \tilde{a}_1 & \ldots & \tilde{a}_m \end{array}: \tilde{y}_1, \ldots, \tilde{y}_d \right\}.
\]
Naturally, $R(x)$ is a function of the characteristic data and the $y_i$, hence, we write
\[
R(x) = \tilde{Y}_0(x)Y_0(x)^{-1}= \tilde{Y}_\infty(x)Y_\infty(x)^{-1} = R(x; y_i, \tilde{y}_i, \kappa_j,\tilde{\kappa}_j,\tilde{\lambda}_l, \lambda_l,
a_n,\tilde{a}_n).
\]
An alternate characterization of \eqref{assoctilde} is the compatibility condition resulting the consistency of~\eqref{tildeev} with~\eqref{qx}, which gives
\begin{gather}\label{comp}
\tilde{A}(x) R(x) = R(qx)A(x),
\end{gather}
which has appeared many times in the literature \cite{Sakai:qP6, Gramani:Isomonodromic, Sakai:Garnier}. Given a suitable parameterization, we obtain a rational map on the co-ordinate system for $A(x)$:
\[
T : \ M_A \to M_{\tilde{A}},
\]
For each $i = 1, \ldots, d$, we use \eqref{comp} to f\/ind a rational mapping $\phi_i$ such that
\[
\tilde{y}_i = \phi_i(M_{A}),
\]
where $M_{A}$ includes all of the variables in \eqref{defining}.

We now draw further correspondence with the framework of Jimbo and Miwa~\cite{Jimbo:Monodromy2}. Let $\{\mu_1, \ldots, \mu_K\}$ be a collection of elements in the characteristic data, and $\{m_1, \ldots, m_K\}$ be a set of integers, then we def\/ine the transformation
\[
T_{\mu_1^{m_1}, \ldots, \mu_K^{m_K}} : \ M_{A} \to M_{\tilde{A}},
\]
to be the transformation that multiplies $\mu_i$ by $q^{m_i}$ leaving all other characteristic data f\/ixed. The~$m_i$ have to be chosen so that~$M_{\tilde{A}}$ satisf\/ies~\eqref{constraint}. The group of Schlesinger transforma\-tions,~$G_A$, is the set of these transformation with the operation of composition
\[
G_A = \langle T_{\mu_1^{m_1}, \ldots, \mu_K^{m_k}}, \circ \rangle.
\]
This group is generated by a set of elementary Schlesinger transformations which only alter two variables, i.e., $K=2$. For example, if wish to multiply $\kappa_1$ by $q$ and $a_1$ by $q^{-1}$, so to preserve~\eqref{constraint}, this would be the elementary transformation $T_{\kappa_1, a_1^{-1}}$. Since the subscripts def\/ine the change in characteristic data, we need only specify the relation between the $y_i$ and $\tilde{y}_i$, hence, we will write
\[
T_{\mu_1^{m_1}, \ldots, \mu_K^{m_K}} : \    \tilde{y}_1 = \phi_1(M_A), \quad
\dots, \quad
\tilde{y}_d = \phi_d(M_A).
\]
We will denote matrix that induces the transformation $T_{\mu_1^{m_1}, \ldots, \mu_K^{m_K}} $ by $R_{\mu_1^{m_1}, \ldots, \mu_K^{m_K}}$ so that the transformation $M_A \to M_{\tilde{A}}$ is specif\/ied by the transformation of the linear problem given by
\[
\tilde{Y}(x) = R_{\mu_1^{m_1}, \ldots, \mu_K^{m_K}} (x) Y(x).
\]
The goal of the the later sections will be to construct a set of elementary Schlesinger transformations that generate $G_A$ and describe the corresponding $R$ matrices.

\section[$q$-Painlev\'e equations]{$\boldsymbol{q}$-Painlev\'e equations}\label{section3}

\begin{figure}[!ht]
\centering
\includegraphics{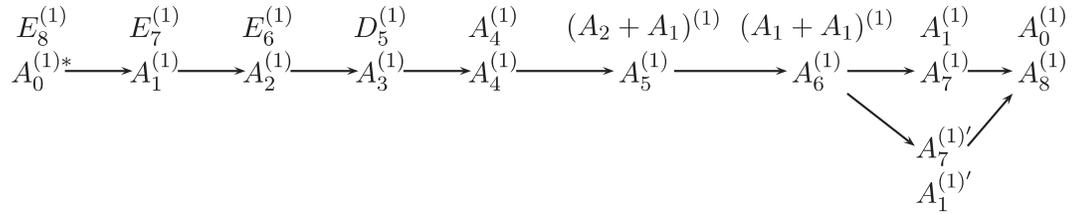}
\caption{This diagram represents the degeneration diagram for the surfaces of initial conditions of multiplicative type~\cite{Sakai:Rational}.\label{degen}}
\end{figure}

We now look at the group of Schlesinger transformations for the $q$-Painlev\'e equations. In accordance with the classif\/ication of Sakai~\cite{Sakai:Rational}, we have ten surfaces considered to be of multiplicative type. Each surface is associated with a root subsystem of a root system of type $E_8^{(1)}$. We may label the surface in two dif\/ferent ways; either the type of the root system of $E_8^{(1)}$, $R$, or the type of the root subsystem of the orthogonal complement of~$R$, $R^{\bot}$. Given a surface of initial conditions associated with a root subsystem of type $R$, the Painlev\'e equation itself admits a~representation of an af\/f\/ine Weyl group of type~$R^{\bot}$ as a group of B\"acklund transformations. The degeneration diagram for these surfaces is shown in Fig.~\ref{degen} where we list both $R$ along with~$R^{\bot}$.

In this article we will discuss the group of connection preserving deformations for $q$-Painlev\'e equations for cases up to and including the $A_3^{(1)}$. As the $A_8^{(1)}$ surface does not correspond to any $q$-Painlev\'e equation, this gives us six cases to consider here. These are also the set of cases in which~$A(x)$ is quadratic in $x$. We will consider the higher degree cases, which include associated linear problem for the Painlev\'e equations whose B\"acklund transformations represent the exceptional af\/f\/ine Weyl groups in a separate article.

\begin{table}[!ht]\centering
\caption{The correspondence between the data that def\/ines the linear problem and the $q$-Painlev\'e equations. This data includes the determinant, and the asymptotic behavior of the solutions at $x= 0$ and $x = \infty$ in~\eqref{fundsols}.}

\vspace{1mm}

\begin{tabular}{ |c | c | c | c | c |}
\hline
 &\tsep{1pt}\bsep{1pt} $\det A$ & $(\mu_1,\mu_2)$ & $(\nu_1,\nu_2)$ & $\dim G_A$ \\ \hline \hline
$\mathrm{P}\big(A_7^{(1)}\big)$\tsep{3pt} & $\kappa_1\kappa_2x^3$ & $(2,1)$ & $(0,3)$ & 3\\
$\mathrm{P}\big(A_7^{(1)'}\big)$ & $\kappa_1\kappa_2x^2$ & $(2,0)$ & $(0,2)$ & 3\\
$\mathrm{P}\big(A_6^{(1)}\big)$ & $\kappa_1\kappa_2x^2(x-a_1)$ & $(2,1)$ & $(0,2)$ & 4 \\
$\mathrm{P}\big(A_5^{(1)}\big)$ & $\kappa_1\kappa_2x(x-a_1)(x-a_2)$ & $(2,1)$ & $(0,1)$ & 5\\
$\mathrm{P}\big(A_4^{(1)}\big)$ & $\kappa_1\kappa_2(x-a_1)(x-a_2)(x-a_3)$ & $(2,1)$ & $(0,0)$ &6 \\
$\mathrm{P}\big(A_3^{(1)}\big)$ & $\kappa_1\kappa_2(x-a_1)(x-a_2)(x-a_3)(x-a_4)$ & $(2,2)$ & $(0,0)$ & 7 \\  \hline
\end{tabular}
\end{table}

It has been well established that the degree one case of \eqref{qx}, where $A(x) = A_0 + A_1x$ is completely solvable in general in terms of hypergeometric functions~\cite{LeCaine}. It is also interesting to note that the basic hypergeometric dif\/ference equation f\/its into the framework of connection preserving deformations. The result of the above transformation group reproduces many known transformations~\cite{GasperRahman}.

\subsection[$q$-$\mathrm{P}\big(A_7^{(1)}\big)$]{$\boldsymbol{q}$-$\boldsymbol{\mathrm{P}\big(A_7^{(1)}\big)}$}

Let us consider the simplest case in a more precise manner than the others as a test case. The simplest $q$-dif\/ference case is the
\[
\left\{ b_0,b_1 : f, g\right\} \to \left\{ \dfrac{b_0}{q}  , b_1q : \tilde{f}, \tilde{g} \right\},
\]
where $q = b_0b_1$ and the evolution is
\begin{gather}\label{qPI}
\tilde{f}f  = \dfrac{b_1(1 - \tilde{g})}{\tilde{g}},\qquad
\tilde{g}g  = \dfrac{1}{f},
\end{gather}
which is also known as $q$-$\mathrm{P}_{\mathrm{I}}$. We note that this transformation gives us a copy of $\mathbb{Z}$ inside the group of B\"acklund transformations \cite{Sakai:Rational}. We note that there exists also exists a Dynkin diagram automorphism in the group of B\"acklund transformations, however, we still do not know how these Dynkin diagram automorphisms manifest themselves in this theory. The associated linear problem, specif\/ied by Murata~\cite{Murata2009}, is given by
\[
Y(qx) = A(x)Y(x) ,
\]
where
\begin{gather*}
A(x) = \begin{pmatrix}\left((x-y) (x-\alpha )+z_1\right) \kappa _1 & \kappa_2 w (x-y) \vspace{1mm}\\
 \dfrac{(x \gamma +\delta ) \kappa _1}{w} & \left(x-y+z_2\right) \kappa _2
 \end{pmatrix}.
\end{gather*}
We f\/ix the determinant
\[
\det A(x) = \kappa_1\kappa_2 x^3,
\]
and let $A(0)$ possess one non-zero eigenvalue,~$\lambda_1$. This allows us to specify the values
\begin{gather*}
\alpha  = \dfrac{\lambda_1 -z_1 \kappa _1+\left(y-z_2\right) \kappa _2}{y \kappa _1},\qquad
\gamma  = \left(-2 y-\alpha +z_2\right) \kappa _2,\qquad
\delta  = \dfrac{\left(y \alpha +z_1\right) \left(y-z_2\right) \kappa _2}{y}.
\end{gather*}
We make a choice of parameterization of $z_1$ and $z_2$, given by
\begin{gather*}
z_1  = \dfrac{y^2}{z},\qquad
z_2  = yz.
\end{gather*}
We will make a full correspondence between $y$ and $z$ with $f$ and $g$. To do this, let us f\/irst consider the full space of Schlesinger transformations on this space.

The problem possesses formal solutions around $x= 0$ and $x= \infty$ specif\/ied by
\begin{gather*}
Y_0(x)  = \left(\begin{pmatrix} -1 & -w y \kappa _2 \vspace{1mm}\\
 \dfrac{z-1}{w} & y (z-1) \kappa _2-\lambda   \end{pmatrix} + O\left(x\right)\right) \begin{pmatrix} e_{q,\lambda_1}(x) & 0 \\ 0 & \dfrac{e_{q,\lambda_2}(x)}{\theta_q\left(x\right)^3} \end{pmatrix},\\
Y_\infty(x)  = \left( I + \dfrac{1}{x}\begin{pmatrix}   \dfrac{q (y+\alpha )}{q-1} & -\dfrac{w \kappa _2}{\kappa _1} \vspace{1mm}\\
 \dfrac{q \gamma }{w} & -\dfrac{q (y+\alpha )}{q-1} \end{pmatrix} + O\left(\frac{1}{x^2} \right) \right)\begin{pmatrix}\dfrac{e_{q,\kappa_1}(x)}{ \theta_q\left(x\right)^2} & 0 \\ 0 & \dfrac{e_{q,\kappa_2}(x)}{\theta_q\left(x\right)} \end{pmatrix}.
\end{gather*}
The asymptotics of these solutions and the zeros of the determinant def\/ine the characteristic data to be
\[
M = \left\{ \begin{array}{c c} \kappa_1 & \kappa_2 \\ \lambda_1 & \lambda_2 \end{array} \right\} ,
\]
meaning our co-ordinate system for the $A$ matrices should be
\[
M_A = \left\{ \begin{array}{c c} \kappa_1 & \kappa_2 \\ \lambda_1 & \lambda_2 \end{array} : w, y, z\right\} ,
\]
where we have the constraint
\[
\lambda_1\lambda_2 = \kappa_1\kappa_2.
\]
This specif\/ies four parameters with one constraint, however, it is easy to verify on the level of parameters (and much more work to verify on the level of the Painlev\'e variables) that
\[
T_{\kappa_1,\lambda_2} \circ T_{\kappa_1,\lambda_1}^{-1} \circ T_{\kappa_2, \lambda_1} = T_{\kappa_1,\lambda_2},
\]
hence, we may regard $T_{\kappa_1,\lambda_2}$ as an element of the group generated by the other three elements. Upon further examination, the action of $R_{\kappa_1,\kappa_2,\lambda_1,\lambda_2}(x) = xI$ is represented by an identity on the Painlev\'e parameters, however, there is a $q$-shift of the asymptotic behaviors at $0$ and $\infty$. We regard this as a trivial action, $T_{\kappa_1,\kappa_2,\lambda_1,\lambda_2}$, whereby, we have the relation
\[
T_{\kappa_1,\lambda_2} \circ T_{\kappa_2,\lambda_1} = T_{\kappa_1,\kappa_2,\lambda_1,\lambda_2},
\]
hence, if we include $T_{\kappa_1,\kappa_2,\lambda_1,\lambda_2}$, if we include $T_{\kappa_1,\kappa_2,\lambda_1,\lambda_2}$ in our generators, we need only consider the two non-trivial translations $T_{\kappa_1,\lambda_1}$ and $T_{\kappa_1,\lambda_2}$. We now proceed to calculate
\begin{gather*}
T_{\kappa_1,\lambda_1} : \ \left\{ \begin{array}{c c} \kappa_1 & \kappa_2 \\ \lambda_1 & \lambda_2 \end{array} : w,y,z \right\}  \to \left\{ \begin{array}{c c} q\kappa_1 & \kappa_2 \\ q \lambda_1 & \lambda_2 \end{array} : \tilde{w},\tilde{y},\tilde{z} \right\},\\
T_{\kappa_1,\lambda_2} : \ \left\{ \begin{array}{c c} \kappa_1 & \kappa_2 \\ \lambda_1 & \lambda_2 \end{array} : w,y,z \right\}  \to \left\{ \begin{array}{c c} q\kappa_1 & \kappa_2 \\ \lambda_1 & q\lambda_2 \end{array} : \tilde{w},\tilde{y},\tilde{z} \right\} ,
\end{gather*}
where in these two cases, the action $(w, y,z) \to (\tilde w, \tilde y, \tilde z)$, is obtained from the left action of a~Schlesinger matrix, $R = R_{\kappa_1,\lambda_1}(x)$ and $R = R_{\kappa_1,\lambda_2}(x)$ respectively. One of the aspects of \cite{studyqPV} was that $R(x)$ may be obtained directly from the solutions via expansions of
\[
R(x) = \tilde{Y}(x) Y(x)^{-1},
\]
where $Y = Y_0$ or $Y = Y_\infty$. Using $Y = Y_0$ an expansion of $R(x)$ around $x = 0$ gives
\[
R(x) = R_0 + O(x) ,
\]
where as using $Y = Y_{\infty}$ an expansion of $R$ around $x= \infty$ gives
\[
R(x) = \begin{pmatrix} x & 0 \\ 0 & 0 \end{pmatrix} + R_0 + O\left( \dfrac{1}{x} \right).
\]
The equality of these gives us that $R(x)$ is linear in $x$, in fact, expanding $R$ around $x = \infty$ to the constant term gives
\begin{gather}\label{Rk1l1}
R(x) = \begin{pmatrix}
 x+\dfrac{q (\tilde{y}-y+\tilde{\alpha}-\alpha)}{q-1} & \dfrac{\kappa_2 w}{\kappa _1} \vspace{1mm}\\
 \dfrac{q \tilde{\gamma}}{\tilde{w}} & 1
\end{pmatrix}.
\end{gather}
This parameterization is the same for $R_{\kappa_1,\lambda_1}$ and $R_{\kappa_1,\lambda_2}$, however, $\tilde{y}$ and~$\tilde{w}$ and hence~$\tilde{\gamma}$ and~$\tilde{\alpha}$ are dif\/ferent for each case. Computing the compatibility condition for the two non-trivial generators gives the following relations
\begin{gather*}
T_{\kappa_1,\lambda_1} : \
\tilde{w} = w \left(\dfrac{q\lambda_1(\tilde{z}-1)}{y\tilde{z}\kappa_1} \right), \qquad
\tilde{y} = \dfrac{q y \lambda  \tilde{z}}{(\tilde{z}-1) \left(q \lambda +y \kappa _2 \tilde{z}\right)},\qquad
\tilde{z} = \dfrac{q z \lambda }{q y^2 \kappa _1-y z \kappa _2},
\\
T_{\kappa_1,\lambda_2} :  \
\tilde{w} = w \left(\dfrac{\kappa_2}{\kappa_1} - \dfrac{qy}{z} \right) ,\qquad
\tilde{y} = \dfrac{\lambda_1}{q\kappa_1y - \kappa_2z}, \qquad
\tilde{z} = \dfrac{q\kappa_1y}{\kappa_2z}.
\end{gather*}
We now note
\begin{gather*}
\mathbb{Z}^3   \cong \langle T_{\kappa_1,\lambda_1}, T_{\kappa_1,\lambda_2}, T_{\kappa_1,\kappa_2,\lambda_1,\lambda_2}\rangle = G_A,
\end{gather*}
which is the lattice of connection preserving deformations.

We now specify the connection preserving deformation that def\/ines $q$-$\mathrm{P}_{\rm I}$ as
\[
q\textrm{-}\mathrm{P}_{\rm I} : \left\{ \begin{array}{c c} \kappa_1 & \kappa_2 \\ \lambda_1 & \lambda_2 \end{array} : w,y,z \right\} \to
\left\{ \begin{array}{c c} \kappa_1 & \kappa_2 \\ q\lambda_1 & \dfrac{\lambda_2}{q} \end{array} : \tilde{w},\tilde{y},\tilde{z} \right\}.
\]
which is decomposed elementary Schlesinger transformations $q$-$\mathrm{P}_{\rm I} = T_{\kappa_1,\lambda_1} \circ T_{\kappa_1,\lambda_2}^{-1}$. A simple calculation reveals
\[
T_{\lambda_1,\lambda_2^{-1}}= T_{\kappa_1,\lambda_1} \circ T_{\kappa_1,\lambda_2}^{-1} : \
\tilde{w} = w(1-\tilde{z}), \qquad
\tilde{y}y = \dfrac{\lambda_1 \tilde{z}}{\kappa_1(\tilde{z}-1)},\qquad
\tilde{z} z = \dfrac{q\kappa_1y}{\kappa_2}.
\]
We make the correspondence with \eqref{qPI} by letting $f = \dfrac{\kappa_2}{q\kappa_1 y}$ and $z= g$, $b_1 = -\kappa_2^2/(q^2\lambda_1\kappa_1)$ and $b_0b_1 = q$. In this way, the one and only translational element of $A_1^{(1)}$ is identif\/ied with $T_{\lambda_1,\lambda_2^{-1}}$.

\subsection[$q$-$\mathrm{P}\big(A_7^{(1)'}\big)$]{$\boldsymbol{q}$-$\boldsymbol{\mathrm{P}\big(A_7^{(1)'}\big)}$}

We note that there are many ways to interpret $q$-$\mathrm{P}\big(A_7^{(1)'}\big)$. We choose but one element of inf\/inite order that we are able to make a correspondence with, for this, in Sakai's notation, we choose $s_1 \circ \sigma_{(1357)(2460)}$, which has the following ef\/fect
\[
\left\{\begin{array}{c} b_0 \,\, b_1 \\ b_2 \end{array} : f, g\right\} \to \left\{\begin{array}{c} \dfrac{b_0}{q} \,\, \dfrac{q}{b_1} \\ b_2b_1^2 \end{array} : \tilde{f}, \tilde{g}\right\},
\]
where
\begin{gather*}
\tilde{f}f = \frac{q+\tilde{g}b_0}{\tilde{g}(\tilde{g}+1)b_0},\qquad
\tilde{g}g = \frac{q}{b_0b_1f^2},
\end{gather*}
which is also known as $q$-$\mathrm{P}_{\mathrm{I}}'$. The surface corresponds to the same af\/f\/ine Weyl group as before, however, the technical dif\/ference is that it corresponds to a copy of a root subsystem of $E_8^{(1)}$ where the lengths of the roots are dif\/ferent from those that correspond to $q$-$\mathrm{P}\big(A_7^{(1)}\big)$  \cite{Sakai:Rational}.

The associated linear problem for $q$-$\mathrm{P}\big(A_7^{(1)'}\big)$, has been given by Murata~\cite{Murata2009}. In terms of the theory presented above, the same theory applies in that the characteristic data is well def\/ined, and the deformations are all prescribed in the same manner as for  $q$-$\mathrm{P}\big(A_7^{(1)}\big)$. In short, we expect a three dimensional lattice of deformations as above. We take
\[
A(x) = \begin{pmatrix} \kappa_1((x-y)(x-\alpha)+z_1) & \kappa_2 w(x-y)\vspace{1mm} \\ \dfrac{\kappa_1}{w}(\gamma x+\delta) & \kappa_2 z_2 \end{pmatrix},
\]
where
\begin{gather*}
\alpha  = \dfrac{-z_1 \kappa _1-z_2 \kappa _2+\lambda _1}{y \kappa _1},\qquad
\gamma  = z_2 \kappa _2-\kappa _2,\qquad
\delta  = y \gamma -y z_2 \kappa _2-\alpha  z_2 \kappa _2,
\end{gather*}
and the choice of $z$ is specif\/ied by
\begin{gather*}
z_1 = \dfrac{y^2}{z}, \qquad
z_2 = z.
\end{gather*}
We f\/ind the solutions are given by
\begin{gather*}
Y_0(x) =  \left(\begin{pmatrix}  -\dfrac{w y (1-z)}{z \kappa _2} & \dfrac{w y \left(y (z-1) \kappa _2-\lambda \right)}{\lambda _1-z \kappa _2} \vspace{1mm}\\
 1-z & y (z-1) \kappa _2-\lambda \end{pmatrix} + O(x)  \right)  \begin{pmatrix} e_{q,\lambda_1}(x) & 0\vspace{1mm}\\ 0 & \dfrac{e_{q,\lambda_2}(x)}{\theta_q\left(x\right)^2 } \end{pmatrix},\\
Y_\infty(x) =  \left( I + \frac{1}{x}\begin{pmatrix} \dfrac{q (y+\alpha )}{q-1} & -\dfrac{w}{\kappa _1} \vspace{1mm}\\
 \dfrac{q \gamma }{w} & -\dfrac{q \left((z-1) \kappa _1 y^2+z \left(\lambda _1-z \kappa _2\right)\right)}{(q-1) y z \kappa _1}\end{pmatrix} + O\left( \dfrac{1}{x^2}\right) \right) \\
\phantom{Y_\infty(x) =}{}  \times \begin{pmatrix} \dfrac{e_{q,\kappa_1}(x)}{\theta_q\left(x\right)^2} & 0\vspace{1mm}\\ 0 & e_{q,\kappa_2}(x) \end{pmatrix},
\end{gather*}
where
\[
\kappa_1\kappa_2 = \lambda_1\lambda_2.
\]
This allows us to specify the characteristic data to be
\[
M = \left\{ \begin{array}{c c} \kappa_1 & \kappa_2 \\ \lambda_1 & \lambda_2 \end{array} \right\},
\]
and the def\/ining co-ordinate system for $A(x)$ to be
\[
M_A = \left\{ \begin{array}{c c} \kappa_1 & \kappa_2 \\ \lambda_1 & \lambda_2 \end{array} : w, y, z\right\}.
\]
For the same reasons as for the previous subsection, it suf\/f\/ices to def\/ine $T_{\kappa_1,\lambda_1}$,  $T_{\kappa_1,\lambda_2}$ and $T_{\kappa_1,\kappa_2,\lambda_1,\lambda_2}$, which is a trivial shift induced by $R_{\kappa_1,\kappa_2,\lambda_1,\lambda_2}(x) = xI$. The matrices $R_{\kappa_1,\lambda_1}(x)$ and $R_{\kappa_1,\lambda_2}(x)$ are given by \eqref{Rk1l1}, allowing us compute the following:
\begin{gather*}
T_{\kappa_1,\lambda_1}  : \
\tilde{w} = \dfrac{q w \lambda _1 (\tilde{z}-1)}{y \kappa _1 \tilde{z}},\qquad
\tilde{y} = \dfrac{y \tilde{z} \left(q \lambda _1-\tilde{z} \kappa _2\right)}{q (\tilde{z}-1) \lambda _1},\qquad
\tilde{z} = \dfrac{z \lambda _1}{y^2 \kappa _1},
\\
T_{\kappa_1,\lambda_2}  : \
\tilde{w} = -qw\dfrac{y}{z},\qquad
\tilde{y}y = \dfrac{\lambda_1}{q\kappa_1},\qquad
\tilde{z}z = \dfrac{\lambda_1}{q\kappa_2}.
\end{gather*}
In a similar manner to the associated linear problem for ${\rm P}\big(A_7^{(1)}\big)$, we have
\begin{gather*}
\mathbb{Z}^3  \cong \langle T_{\kappa_1,\lambda_1}, T_{\kappa_1,\lambda_2}, T_{\kappa_1,\kappa_2,\lambda_1,\lambda_2} \rangle =  G_A ,
\end{gather*}
forming the lattice of connection preserving deformations.

We now identify the evolution of $q$-$\mathrm{P}_{\rm I}'$ as a decomposition into the basis for elementary Schlesinger transformations above.
$q$-$\mathrm{P}_{\rm I}' = T_{\kappa_1,\lambda_1} \circ T_{\kappa_1,\lambda_2}^{-1}$. The action is specif\/ied by
\[
T_{\kappa_1,\lambda_1} \circ T_{\kappa_1,\lambda_2}^{-1} :
\
\tilde{w} = w(1-\tilde{z}), \qquad
\tilde{y}y = \dfrac{\tilde{z} \left(q \lambda _1-\kappa _2 \tilde{z}\right)}{q \kappa _1 (\tilde{z}-1)},\qquad
\tilde{z}z = \dfrac{q\kappa_1}{\kappa_2} y^2,
\]
where we may identify with the above evolutions as $z = -1/g $, $y = \sqrt{b_0b_1}f$ and $\kappa_2b_0 = q^2\lambda_1- q^2\kappa_1$.

\subsection[$q$-$\mathrm{P}\big(A_6^{(1)}\big)$]{$\boldsymbol{q}$-$\boldsymbol{\mathrm{P}\big(A_6^{(1)}\big)}$}

We now increase the dimension of the underlying lattice by introducing non-zero root of $\det A$. We note that in accordance with the notation of Sakai \cite{Sakai:Rational}, we choose $\sigma \circ \sigma$ to def\/ine the evolution of $q$-$\mathrm{P}_{\rm II}$. This system may be equivalently written as
\[
\left\{\begin{array}{c} b_0 \,\, b_1 \\ b_2 \end{array} : f, g\right\} \to \left\{\begin{array}{c} b_0 \,\, b_1 \\ qb_2 \end{array} : \tilde{f}, \tilde{g}\right\},
\]
where $q = b_0b_1$ and
\begin{gather}\label{qPII}
\tilde{f}f = -\dfrac{b_2 \tilde{g}}{\tilde{g}+b_1b_2},\qquad
\tilde{g}g = b_1b_2(1-f).
\end{gather}
We f\/ind a parameterization of the associated linear problem to be given by
\[
A(x) = \begin{pmatrix} \kappa_1((x-\alpha)(x-y)+z_1) & \kappa_2 w(x-y) \vspace{1mm}\\
\dfrac{\kappa_1}{w}(\gamma x + \delta) & \kappa_2(x-y + z_2) \end{pmatrix},
\]
where by letting
\[
\det A(x) = \kappa_1\kappa_2x^2(x-a_1),
\]
we readily f\/ind the entries of $A(x)$ are given by
\begin{gather*}
\alpha  =\dfrac{-z_1 \kappa _1+\left(y-z_2\right) \kappa _2+\lambda _1}{y \kappa _1},\qquad
\gamma  =-2 y-\alpha +a_1+z_2,\qquad
\delta  =\dfrac{\left(y \alpha +z_1\right) \left(y-z_2\right)}{y}.
\end{gather*}
We take a choice of $z$ to be specif\/ied by
\begin{gather*}
z_1 =\dfrac{y \left(y-a_1\right)}{z},\qquad
z_2 =y z.
\end{gather*}
The formal symbolic solutions are specif\/ied by
\begin{gather*}
Y_0(x) =  \left( \begin{pmatrix}  -\dfrac{w (1-z)}{z-1} & -w y \kappa _2 \vspace{1mm}\\
 1-z & y (z-1) \kappa _2-\lambda _1 \end{pmatrix} + O(x) \right) \begin{pmatrix} e_{q,\lambda_1}(x) & 0 \vspace{1mm}\\ 0 & \dfrac{e_{q,\lambda_2}(x)}{\theta_q\left(x\right)^2}  \end{pmatrix},\\
Y_\infty(x) =  \left(I + \dfrac{1}{x}\begin{pmatrix}  \dfrac{q (y+\alpha )}{q-1} & -\dfrac{w \kappa _2}{\kappa _1} \vspace{1mm}\\
 \dfrac{q \gamma }{w} & -\dfrac{q y (z-1) \left(\left(y-a_1\right) \kappa _1-z \kappa _2\right)+q z \lambda _1}{(q-1) y z \kappa _1} \end{pmatrix} + O\left(\dfrac{1}{x^2}\right) \right) \\
  \phantom{Y_\infty(x) =}{}
  \times \begin{pmatrix}\dfrac{e_{q,\kappa_1}(x)}{ \theta_q\left(x\right)^2} & 0 \vspace{1mm}\\ 0 & \dfrac{e_{q,\kappa_2}(x)}{\theta_q\left(x\right)} \end{pmatrix}
\end{gather*}
and the zeros of $\det A(x)$ def\/ine the characteristic data to be
\[
M = \left\{ \begin{array}{c c c} \kappa_1 & \kappa_2 & \\ \lambda_1 & \lambda_2 & a_1 \end{array} \right\}.
\]
This allows us to write co-ordinate system for $A$ as
\[
M_A = \left\{ \begin{array}{c c c} \kappa_1 & \kappa_2 & \\ \lambda_1 & \lambda_2 & a_1 \end{array} : w, y, z\right\},
\]
where we have the constraint
\[
\lambda_1\lambda_2 = -a_1\kappa_1\kappa_2.
\]
For the same reasons as previous subsections, it suf\/f\/ices to def\/ine $T_{\kappa_1,\lambda_1}$, $T_{\kappa_1,\lambda_1}$ and $T_{\kappa_1,\kappa_2,\lambda_1,\lambda_2}$, however, we also need to choose an element that changes $a_1$. We choose $T_{a_1,\lambda_1}$, which is induced by some $R_{a_1,\lambda_1}$. By solving the scalar determinantal problem, one arrives at the conclusion
\[
\det R_{a_1,\lambda_1} = \dfrac{x}{x-qa_1}.
\]
When we consider the change in $\kappa_i$ values, it is clear that by expanding $R_{a_1,\lambda_1}$ around $x= \infty$ using $R_{a_1,\lambda_1} = \tilde{Y}_\infty Y_\infty^{-1}$ we have the expansion
\[
R_{a_1,\lambda_1} = I + \dfrac{1}{x} \begin{pmatrix}
 \dfrac{q (y+\alpha )}{q-1} & -\dfrac{w \kappa _2}{\kappa _1} \vspace{1mm}\\
 \dfrac{q \gamma }{w} & -\dfrac{q \left(y+\alpha -a_1\right)}{q-1}
\end{pmatrix} + O\left(\dfrac{1}{x^2} \right).
\]
This means that we may deduce that $R_{a_1,\lambda_1}$ is given by
\begin{gather}\label{Ra1l1}
R_{a_1,\lambda_1} = \dfrac{1}{x-qa_1} \begin{pmatrix}
x-\dfrac{q (y-\tilde{y}+\alpha-\tilde{\alpha})}{q-1}-q a_1& \dfrac{\kappa_2(w-\tilde{w})}{\kappa _1} \vspace{1mm}\\
 \dfrac{q (\tilde{w} \gamma - w \tilde{\gamma} )}{w \tilde{w}} & x+\dfrac{q (y-\tilde{y}+\alpha -\tilde{\alpha})}{q-1}
\end{pmatrix}.
\end{gather}
We can take $R_{\kappa_1,\lambda_1}$ and $R_{\kappa_1,\lambda_2}$ to be of the form of \eqref{Rk1l1}. Computing the compatibility reveals
\begin{gather*}
T_{a_1,\lambda_1} : \
\tilde{w} = w(1-\tilde{z}),\qquad
\tilde{y}y=\dfrac{\tilde{z} \lambda _1}{(\tilde{z}-1) \kappa _1},\qquad
\tilde{z}z=\dfrac{q \left(y-a_1\right) \kappa _1}{\kappa _2},
\\
T_{\kappa_1,\lambda_1}: \
\tilde{w} = \dfrac{w \left(q \lambda _1-\kappa _2 \tilde{y} (\tilde{z}-1)\right)}{\kappa _1 \tilde{y}},\qquad
\tilde{y}y = \dfrac{z\lambda_1}{\kappa_1(\tilde{z}-1)},\qquad
\tilde{z} = \dfrac{q y a_1 \kappa _1+q z \lambda _1}{q y^2 \kappa _1-y z \kappa _2},
\\
T_{\kappa_1,\lambda_2} : \
\tilde{w} = \dfrac{w \kappa _2 (1-\tilde{z})}{\kappa _1},\qquad
\tilde{y}y = \dfrac{\lambda _1 \tilde{z}}{q \kappa _1 (\tilde{z}-1)},\qquad
\tilde{z}z = \dfrac{q \left(y-a_1\right) \kappa _1}{\kappa _2}.
\end{gather*}
The full group of Schlesinger transformations is given by
\[
\mathbb{Z}^4 \cong \langle T_{a_1,\lambda_1}, T_{\kappa_1,\lambda_1} , T_{\kappa_1,\lambda_2},  T_{\kappa_1,\kappa_2,\lambda_1,\lambda_2} \rangle = G_A,
\]
which is the lattice of connection preserving deformations.

To obtain a full correspondence of $T_{a_1,\lambda_1}$ with \eqref{qPII} we let
\begin{gather*}
y  = a_1f,\qquad
z  = -g\dfrac{b_0}{b_2},
\end{gather*}
with the relations $b_0 = \dfrac{\lambda_1}{a_1\kappa_2}$ and $b_2 = -\dfrac{\lambda_1}{q a_1^2\kappa_1}$. We know that the action of $T_{a_1,\lambda_1} \circ T_{\kappa_1,\lambda_2^{-1}}$ is trivial. I.e., we have
\[
T_{a_1,\lambda_1} \circ T_{\kappa_1,\lambda_2^{-1}} : \  \tilde{w} = \dfrac{\kappa_1w}{\kappa_2},\qquad \tilde{y} = q y, \qquad \tilde{z} = z,
\]
which leaves $f$ and $g$ invariant. When we restrict our attention to $f$ and $g$ alone we only have a two dimensional lattice of transformations, corresponding to $(A_1+A_1)^{(1)}$.

\subsection[$q$-$\mathrm{P}\big(A_5^{(1)}\big)$]{$\boldsymbol{q}$-$\boldsymbol{\mathrm{P}\big(A_5^{(1)}\big)}$}

As a natural progression from the above cases, we allow two non-zero roots of the determinant. In doing so, we obtain an associated linear problem for both $q$-$\mathrm{P}_{\rm III}$ and $q$-$\mathrm{P}_{\rm IV}$. We also introduce a symmetry as one is able to naturally permute the two roots of the determinant, which we will formalize later. We turn to the particular presentation of Noumi et al.~\cite{Noumi}. We present a version of $q$-$\mathrm{P}_{\rm III}$ as
\[
\left\{ \begin{array}{c} b_0 \,\, b_1 \\ b_2\,\, c \end{array} : f,g \right\} \to \left\{ \begin{array}{c} q b_0 \,\, b_1/q \\ b_2 \,\, c  \end{array} : \tilde{f},\tilde{g} \right\},
\]
where $q = b_0b_1b_2$ and
\begin{gather*}
\tilde{f}  = \dfrac{qc}{fg} \dfrac{1+ gb_0}{g+b_0},\qquad
\tilde{g}  = \dfrac{qc^2}{g\tilde{f}}\dfrac{b_1+ q\tilde{f}}{q+b_1\tilde{f}}.
\end{gather*}
Using the same variables, a representation of $q$-$\mathrm{P}_{\rm IV}$ is given by
\[
\left\{ \begin{array}{c} b_0 \,\, b_1 \\ b_2\,\, c \end{array} : f,g \right\} \to \left\{ \begin{array}{c} b_0 \,\, b_1 \\ b_2 \,\, qc  \end{array} : \tilde{f},\tilde{g} \right\},
\]
where
\begin{gather*}
\tilde{f}f = \dfrac{c^2 q b_1 \left(1+gb_0 +fg b_1 b_0\right) b_2}{q b_1 b_2 c^2+g+f g b_1},\qquad
\tilde{g}g = \dfrac{b_0 b_1 \left(q \left(g b_0+1\right) b_2 c^2+f g\right)}{\left(g b_0 \left(f b_1+1\right)+1\right)}.
\end{gather*}
These two equations are interesting as a pair as they have the same surface of initial conditions. If any natural correspondence was to be sought between the theory of the associated linear problems and the theory of the surfaces of initial conditions \cite{Sakai:Rational}, these two equations should possess the same associated linear problem up to reparameterizations. It is interesting to note that the associated linear problems def\/ined by \cite{Murata2009} do indeed possess the same characteristic data and asymptotics. We unify them by writing them as one linear problem, given by
\[
A(x) = \begin{pmatrix} \kappa_2 ((x-\alpha)(x-y) + z_1) & \kappa_2 w (x- y)\vspace{1mm}\\
\dfrac{\kappa_1}{w}(\gamma x + \delta) & \kappa_2(x-y + z_2) \end{pmatrix},
\]
where
\[
\det A(x) = \kappa_1\kappa_2x(x-a_1)(x-a_2).
\]
This determinant and allowing $A(0)$ to have one non-zero eigenvalue, $\lambda_1$, allows us to specify the parameterization as
\begin{gather*}
\alpha  = \dfrac{-z_1 \kappa _1+\left(y-z_2\right) \kappa _2+\lambda _1}{y \kappa _1},\qquad\!
\gamma  = a_1+a_2+z_2-2 y-\alpha ,\qquad\!
\delta  = \dfrac{\left(y \alpha +z_1\right)\left(y-z_2\right)}{y}.
\end{gather*}
We choose $z$ to be specif\/ied by
\begin{gather*}
z_1  =  \dfrac{y \left(y-a_1\right)}{z},\qquad
z_2  = z \left(y-a_2\right).
\end{gather*}
The formal solutions are given by
\begin{gather*}
Y_0(x) =  \left( \begin{pmatrix}  w y & w y \kappa _2 \vspace{1mm}\\
 y-z y+z a_2 & \left(-z y+y+z a_2\right) \kappa _2+\lambda _1 \end{pmatrix} + O(x) \right)
  \begin{pmatrix} e_{q,\lambda_1}(x) & 0 \vspace{1mm}\\ 0 & \frac{e_{q,\lambda_2}(x)}{\theta_q\left(x\right)} \end{pmatrix},\\
Y_\infty(x) =  \left( I + \dfrac{1}{x}\begin{pmatrix} \dfrac{q (y+\alpha )}{q-1} & -\dfrac{w \kappa _2}{\kappa _1} \vspace{1mm}\\
 \dfrac{q \gamma }{w} & -\dfrac{q \left(y+\alpha -a_1-a_2\right)}{q-1}\end{pmatrix} + O\left(\dfrac{1}{x^2}\right)\! \right)\!
 \begin{pmatrix}\dfrac{e_{q,\kappa_1}(x)}{ \theta_q\left(x\right)^2} & 0 \vspace{1mm}\\ 0 & \dfrac{e_{q,\kappa_2}(x)}{\theta_q\left(x\right)} \end{pmatrix},
\end{gather*}
which is enough to def\/ine
\[
M = \left\{ \begin{array}{c c c c}
\kappa_1 & \kappa_2 & \lambda_1 & \lambda_2 \\
a_1 & a_2 & &
\end{array}\right\},
\]
hence, the co-ordinates for $A(x)$ may be stated as
\[
M_A = \left\{ \begin{array}{c c c c}
\kappa_1 & \kappa_2 & \lambda_1 & \lambda_2 \\
a_1 & a_2 & &
\end{array} : w,y,z\right\}.
\]
We notice that there is a natural symmetry introduced in the parameterization, that is that $A$ is left invariant under the action
\[
s_1 : \left\{ \begin{array}{c c c c} \kappa_1 & \kappa_2 & & \\ \lambda_1 & \lambda_2 & a_1 & a_2 \end{array}: w,y,z \right\} \to \left\{ \begin{array}{c c c c} \kappa_1 & \kappa_2 & & \\ \lambda_1 & \lambda_2 & a_2 & a_1 \end{array}: w,y,z\dfrac{y-a_2}{y-a_1} \right\}.
\]
If we include $s_1$ as a symmetry, we need only specify $T_{a_1,\lambda_1}$, $T_{\kappa_1,\lambda_1}$ and $T_{\kappa_1,\lambda_2}$ as we may exploit this symmetry to obtain $T_{a_2,\lambda_1} = s_1 \circ T_{a_1,\lambda_1} \circ s_1$. The matrix $R_{a_1,\lambda_1}$ is of the form \eqref{Ra1l1} and the matrices $R_{\kappa_1,\lambda_1}$ and $R_{\kappa_1,\lambda_2}$ are of the form \eqref{Rk1l1}. We may use the compatibility conditions to obtain the following correspondences:
\begin{gather*}
T_{a_1,\lambda_1}  : \
\tilde{w} = w (1-\tilde{z}),\qquad
\tilde{y}y = \dfrac{\tilde{z}\left(\tilde{z} a_2 \kappa_2+q \lambda_1\right)}{q(\tilde{z}-1) \kappa _1},\qquad
\tilde{z}z = \dfrac{q y \left(y-a_1\right) \kappa _1}{\left(y-a_2\right) \kappa _2},
\\
T_{\kappa_1,\lambda_1}  :  \
\tilde{w} = \dfrac{w \left(q y \kappa _1-z \kappa _2\right) (\tilde{z}-1)}{z \kappa _1},\qquad
\tilde{y}y = \dfrac{z \left(q \lambda _1+a_2 \kappa _2 \tilde{z}\right)}{q \kappa _1 (\tilde{z}-1)},\qquad
\tilde{z} = \dfrac{q y a_1 \kappa _1+q z \lambda _1}{q y^2 \kappa _1-y z \kappa _2},
\\
T_{\kappa_1,\lambda_2}  : \
\tilde{w} = w \left(\dfrac{q \left(a_1-y\right)}{z}+\dfrac{\kappa _2}{\kappa _1}\right),\qquad
\tilde{y}y = \dfrac{\left(z a_2 \kappa _2+\lambda _1\right) \tilde{z}}{q \kappa _1 (\tilde{z}-1)},\\
\hphantom{T_{\kappa_1,\lambda_2}  : \  }{} \
\tilde{z}z =-\dfrac{q \kappa _1 \left(z a_1 a_2 \kappa _2+\left(a_1-y\right) \lambda _1\right)}{\kappa _2 \left(a_2 \left(z \kappa _2-q y
   \kappa _1\right)+\lambda _1\right)}.
\end{gather*}
We may now specify $T_{a_2,\lambda_1} = s_1 \circ T_{a_1,\lambda_1} \circ s_1$.
\[
T_{a_2,\lambda_1} : \
\tilde{w} = w \left(1-\dfrac{q y \kappa _1}{z \kappa _2}\right),\qquad
\tilde{y} = \dfrac{q y a_1 \kappa _1+q z \lambda _1}{q y z \kappa _1-z^2 \kappa _2},\qquad
\tilde{z}z =\dfrac{q y \kappa _1 \left(\tilde{y}-a_1\right)}{\kappa _2 \left(\tilde{y}-q a_2\right)}.
\]
This gives us the full lattice
\begin{gather*}
 \mathbb{Z}^5  \cong \langle T_{a_1,\lambda_1}, T_{a_2,\lambda_1} , T_{\kappa_1,\lambda_1}, T_{\kappa_1,\lambda_2},  T_{\kappa_1,\kappa_2,\lambda_1,\lambda_2} \rangle = G_A.
\end{gather*}

We identify the evolution of $q$-$\mathrm{P}_{\rm III}$ with $T_{a_1,\lambda_1}$ under the change of variables
\begin{gather*}
y  = \dfrac{qb_0b_1^2\lambda_1}{g\kappa_2},\qquad
z  = -\dfrac{f}{q b_1},
\end{gather*}
so long as the following relations hold:
\begin{gather*}
b_0^2  = \dfrac{a_1}{a_2},\qquad
\lambda_2  = c^2 \kappa_2,\qquad
\kappa_2a_2  = -qb_1^2\lambda_1.
\end{gather*}
The symmetry between $a_1$ and $a_2$ alternates between the two translational components of the lattice of translations.

With the above identif\/ication, we also deduce $q$-$\mathrm{P}_{\rm VI}$ may be identif\/ied with
\begin{gather*}
T_{a_1,\lambda_1} \! \circ T_{a_2,\lambda_1}\! \circ T_{\kappa_1,\lambda_2} \! \circ T_{\kappa_1, \lambda_1}^{-1}\!:
\
\tilde{w}=w-\dfrac{q w \left(y-a_1\right) \kappa _1}{z \kappa_2},\qquad
\tilde{y} = \dfrac{q \left(y-a_1\right) \lambda _1-q z a_1 a_2 \kappa _2}{y \left(q \left(y-a_1\right) \kappa _1-z \kappa _2\right)},\\
\phantom{T_{a_1,\lambda_1} \! \circ T_{a_2,\lambda_1}\! \circ T_{\kappa_1,\lambda_2} \! \circ T_{\kappa_1, \lambda_1}^{-1}\!: \ }{} \
\tilde{z} = -\dfrac{q \kappa _1 \left(z a_1 a_2 \kappa_2+\left(a_1-y\right) \lambda _1\right)}{z \kappa _2 \left(a_2 \left(z \kappa _2-q y \kappa _1\right)+\lambda _1\right)}.
\end{gather*}
We also note that in addition to $T_{\kappa_1,\kappa_2,\lambda_1,\lambda_2}$ being trivial, we have that according to the variab\-les~$f$ and~$g$, the transformation $T_{\kappa_1,\lambda_1^{-1},a_1^{-1},a_2^{-1}} = T_{a_1,\lambda_1}^{-1} \circ T_{\kappa_1,\lambda_1} \circ T_{a_2,\lambda_1}^{-1}$ is also trivial.

\subsection[$q$-$\mathrm{P}\big(A_4^{(1)}\big)$]{$\boldsymbol{q}$-$\boldsymbol{\mathrm{P}\big(A_4^{(1)}\big)}$}

The next logical step in the progression of associated linear problems is to assume that the determinant has three non-zero roots. This brings us to the case of $q$-$\mathrm{P}_{\rm V}$:
\[
\left\{\begin{array}{c} b_0\,\, b_1\,\, b_2 \\ b_3 \,\, b_4 \end{array} : f, g \right\}\to \left\{\begin{array}{c} qb_0\,\, b_1\,\, b_2 \\ b_3 \,\, b_4/q \end{array} : \tilde{f}, \tilde{g} \right\}
\]
where
\begin{gather*}
f \tilde{f}  =\dfrac{b_0+g b_2}{b_0 \left(g b_2+1\right) \left(g b_1 b_2+1\right) b_3},\qquad
g \tilde{g} =\dfrac{(\tilde{f}-1) b_0 \left(\tilde{f} b_3-1\right)}{\tilde{f} b_2 \left(\tilde{f} b_0 b_1 b_2 b_3-1\right)}.
\end{gather*}
The associated linear problem was introduced by Murata \cite{Murata2009}, and the particular group has been explored in connection with the big $q$-Laguerre polynomials~\cite{studyqPV}. This case introduces two new symmetries; the f\/irst is that we have the symmetric group on the three roots of $\det A$, and the other is that the two solutions at $x= 0$ are of the same order. The result is that they are interchangeable on the level of solutions of a single second order $q$-dif\/ference equation. The of\/fshoot of this is, by including extra symmetries, we need only compute two non-trivial connection preserving deformations.

Let us proceed by f\/irst giving a parameterization of the associated linear problem. We let
\[
A(x) = \begin{pmatrix} \kappa_1((x-\alpha)(x-y)+ z_1) & \kappa_2 w (x -y) \vspace{1mm}\\
\dfrac{\kappa_1}{w} (\gamma x+ \delta) & \kappa_2(x-y+z_2)
 \end{pmatrix},
\]
where
\[
\det A(x) = \kappa_1\kappa_2(x-a_1)(x-a_2)(x-a_3).
\]
Dif\/ferently to previous case, the determinant of $A(0)$ is non-zero, hence, we require that the eigenvalues of $A(0)$, $\lambda_1$ and $\lambda_2$, are both not necessarily zero. This determines that the parame\-ters in~$A(x)$ are specif\/ied by
\begin{gather*}
\alpha  =\dfrac{-z_1 \kappa _1+y \kappa _2-z_2 \kappa _2+\lambda _1+\lambda _2}{y \kappa _1},\qquad
\gamma  =-2 y-\alpha +a_1+a_2+a_3+z_2, \\
\delta  = \dfrac{\left(y \alpha +z_1\right) \left(y-z_2\right)-a_1 a_2 a_3}{y},
\end{gather*}
where we choose our parameter $z$ to be specif\/ied by
\begin{gather*}
z_1  = \dfrac{\left(y-a_1\right) \left(y-a_2\right)}{z},\qquad
z_2  = z \left(y-a_3\right).
\end{gather*}
Our constraint is now
\[
\kappa_1\kappa_2a_1a_2a_3 = -\lambda_1 \lambda_2.
\]
The fundamental solutions are of the form
\begin{gather*}
Y_0(x)  = \left( \begin{pmatrix}  -w y & -w y \kappa _2 \vspace{1mm}\\
 y (z-1)-z a_3-\dfrac{\lambda _2}{\kappa _2} & \left(y (z-1)-z a_3\right) \kappa _2-\lambda _1 \end{pmatrix} + O(x) \right) \\
 \phantom{Y_0(x)  =}{}\times
   \begin{pmatrix} e_{q,\lambda_1}(x) & 0 \vspace{1mm}\\ 0 & e_{q,\lambda_2}(x) \end{pmatrix},\\
Y_\infty(x) =  \left(I + \dfrac{1}{x}\begin{pmatrix} \dfrac{q (y+\alpha )}{q-1} & -\dfrac{w \kappa _2}{\kappa _1} \vspace{1mm}\\
 \dfrac{q \gamma }{w} & -\dfrac{q \left(y+\alpha -a_1-a_2-a_3\right)}{q-1}\end{pmatrix} + O\left(\dfrac{1}{x^2}\right) \right)\\
\phantom{Y_\infty(x) =}{}  \times\begin{pmatrix}\dfrac{e_{q,\kappa_1}(x)}{ \theta_q\left(x\right)^2} & 0 \vspace{1mm}\\ 0 & \dfrac{e_{q,\kappa_2}(x)}{\theta_q\left(x\right)} \end{pmatrix}.
\end{gather*}
This specif\/ies that our characteristic data may be taken to be
\[
M = \left\{ \begin{array}{c c c c} a_1 & a_2 & a_3 &  \\ \kappa_1 & \kappa_2 & \lambda_1 & \lambda_2 \end{array} \right\},
\]
and hence, our co-ordinate space for $A(x)$ is given by
\[
M_A = \left\{ \begin{array}{c c c c} a_1 & a_2 & a_3 &  \\ \kappa_1 & \kappa_2 & \lambda_1 & \lambda_2 \end{array} : w, y, z\right\}.
\]
We identify immediately that the parametrization is invariant under the following transformations
\begin{gather*}
s_0  : \ \left\{ \begin{array}{c c c c} a_1 & a_2 & a_3 &  \\ \kappa_1 & \kappa_2 & \lambda_1 & \lambda_2 \end{array} : w, y, z\right\} \to \left\{ \begin{array}{c c c c} a_1 & a_2 & a_3 &  \\ \kappa_1 & \kappa_2 & \lambda_2 & \lambda_1 \end{array} : w, y, z\right\},\\
s_1  : \ \left\{ \begin{array}{c c c c} a_1 & a_2 & a_3 &  \\ \kappa_1 & \kappa_2 & \lambda_1 & \lambda_2 \end{array} : w, y, z\right\} \to \left\{ \begin{array}{c c c c} a_2 & a_1 & a_3 &  \\ \kappa_1 & \kappa_2 & \lambda_1 & \lambda_2 \end{array} : w, y, z\right\},\\
s_2  : \ \left\{ \begin{array}{c c c c} a_1 & a_2 & a_3 &  \\ \kappa_1 & \kappa_2 & \lambda_1 & \lambda_2 \end{array} : w, y, z\right\} \to \left\{ \begin{array}{c c c c} a_1 & a_3 & a_2 &  \\ \kappa_1 & \kappa_2 & \lambda_1 & \lambda_2 \end{array} : w, y, z\dfrac{y-a_3}{y-a_2}\right\}.
\end{gather*}
If we now specify $T_{\kappa_1,\lambda_1}$, the trivial transformation, $T_{\kappa_1,\kappa_2,\lambda_1,\lambda_2}$ and $T_{a_1,\lambda_1}$, we may obtain a~ge\-ne\-rating set using the symmetries to obtain
\begin{gather*}
T_{a_2,\lambda_1}  = s_1 \circ T_{a_1,\lambda_1} \circ s_1,\qquad
T_{a_3,\lambda_1}  = s_2 \circ s_1 \circ T_{a_1,\lambda_1} \circ s_1 \circ s_2,\qquad
T_{\kappa_1,\lambda_2}  = s_0 \circ T_{\kappa_1,\lambda_1} \circ s_0.
\end{gather*}
The transformations $T_{\kappa_1,\lambda_1}$ and $T_{a_1,\lambda_1}$ are specif\/ied by the relations
\begin{gather*}
T_{a_1,\lambda_1}  : \
\tilde{w} =\dfrac{w \left(\left(y-a_1\right) a_2 a_3 \kappa _1 \kappa _2+\left(q y \left(y-a_1\right) \kappa _1+z \left(a_3-y\right) \kappa
   _2\right) \lambda _1\right)}{\kappa _2 \left(\left(y-a_1\right) a_2 a_3 \kappa _1+z \left(a_3-y\right) \lambda _1\right)}, \\
\phantom{T_{a_1,\lambda_1}  : \ }{} \
\tilde{y} =-\dfrac{a_2 (w-\tilde{w}) \left(q \lambda _1+a_3 \kappa _2 \tilde{z}\right)}{q \lambda _1 (\tilde{w}+w (\tilde{z}-1))},\\
\phantom{T_{a_1,\lambda_1}  : \ }{} \
\tilde{z} = \dfrac{z \lambda _1 (w-\tilde{w})+a_2 \kappa _1 \left(w \left(y-a_1\right)+\left(y (z-1)+a_1\right) \tilde{w}\right)}{w
   \left(y-a_1\right) a_2 \kappa _1+w z \lambda _1},
\\
T_{\kappa_1,\lambda_1}  : \
\tilde{w}=\dfrac{w \left(-q \kappa _1 y^2+\left(q \left(a_1+a_2\right) \kappa _1+z \kappa _2\right) y-q a_1 a_2 \kappa _1+q z \lambda
   _1\right)}{y z \kappa _1},\\
\phantom{T_{\kappa_1,\lambda_1}  : \ }{} \
\tilde{y}= \dfrac{q \lambda _1+a_3 \kappa _2 \tilde{z}}{q \psi  \kappa _1 \lambda _1-q \psi  \kappa _1 \lambda _1 \tilde{z}},\qquad
\tilde{z}=\dfrac{q \left(\psi  a_1 \kappa _1-1\right) \left(\psi  a_2 \kappa _1-1\right)}{z \psi  \left(\kappa _2+q \psi  \kappa _1 \lambda_1\right)},
\end{gather*}
where we have introduced
\[
\psi = \dfrac{y}{a_1 a_2 \kappa _1-z \lambda _1},
\]
for convenience. We may now identify the lattice of connection preserving deformations as
\begin{gather*}
\mathbb{Z}^6  \cong \langle T_{\kappa_1,\lambda_1}, T_{\kappa_1,\lambda_2},  T_{\kappa_1,\kappa_2,\lambda_1,\lambda_2}, T_{a_1,\lambda_1}, T_{a_2,\lambda_1}, T_{a_3,\lambda_1} \rangle = G_A.
\end{gather*}
We may identify the evolution of $q$-$\mathrm{P}_{\rm V}$ with the Schlesinger transformation
\begin{gather*}
q\textrm{-}\mathrm{P}_{\rm V} = T_{a_1,a_2,\lambda_1,\lambda_2} : \
\tilde{w} = w(1-\tilde{z}),\qquad
\tilde{y}y = \dfrac{\left(q \lambda _1+a_3 \kappa _2 \tilde{z}\right) \left(\lambda _1 \tilde{z}-q a_1 a_2 \kappa _1\right)}{q \kappa _1 \lambda _1 (\tilde{z}-1)},\\
\phantom{q\textrm{-}\mathrm{P}_{\rm V} = T_{a_1,a_2,\lambda_1,\lambda_2} : \ }{} \
\tilde{z}z = \dfrac{q \left(y-a_1\right) \left(y-a_2\right) \kappa _1}{\left(y-a_3\right) \kappa _2}.
\end{gather*}
where the correspondence between $y$ and $z$ with $f$ and $g$ is given by
\begin{gather*}
y  = \dfrac{a_2}{f},\qquad
z  = -\dfrac{b_2}{b_0}g,
\end{gather*}
under the relations
\begin{gather*}
a_1  = a_2b_3,\qquad
a_2  = a_3b_4,\qquad
qa_2b_0\kappa_1  = - b_2\kappa_2, \qquad
b_0b_1b_4 \lambda_1  = a_2\kappa_2.
\end{gather*}
In addition to $T_{\kappa_1,\kappa_2,\lambda_1,\lambda_2}$ being trivial, we also have that the transformation $T_{a_1,a_2,a_3,\kappa_1^{-1}, \lambda_1,\lambda_2}$ is trivial.

\subsection[$q$-$\mathrm{P}\big(A_3^{(1)}\big)$]{$\boldsymbol{q}$-$\boldsymbol{\mathrm{P}\big(A_3^{(1)}\big)}$}

This particular case was originally the subject of the article by Jimbo and Sakai~\cite{Sakai:qP6}, which introduced the idea of the preservation of a connection matrix~\cite{Birkhoff}. The article was also responsible for introducing a discrete analogue of the sixth Painlev\'e equation \cite{Sakai:qP6}, given by
\[
\left\{ \begin{array}{c c c c}
b_1 & b_2 & b_3 & b_4 \\
b_5 & b_6 & b_7 & b_8
\end{array} : f,g \right\} \to \left\{\begin{array}{c c c c}
qb_1 & qb_2 & b_3 & b_4 \\
qb_5 & qb_6 & b_7 & b_8
\end{array} : \tilde{f},\tilde{g} \right\} ,
\]
where
\begin{gather}\label{qPVI}
\dfrac{\tilde{f} f}{b_7b_8}  = \dfrac{\tilde{g}-qb_1}{\tilde{g}-b_3} \dfrac{\tilde{g}-qb_2}{\tilde{g}-b_4}, \qquad
\dfrac{\tilde{g} g}{b_3b_4}  = \dfrac{f-b_5}{f-b_7}\dfrac{f-b_6}{f-b_8},
\end{gather}
where
\[
q = \dfrac{b_3b_4b_5b_6}{b_1b_2b_7b_8}.
\]
The way in which $q$-$\mathrm{P}_{\rm VI}$, or $q$-$\mathrm{P}\big(A_3^{(1)}\big)$, was derived by Jimbo and Sakai \cite{Sakai:qP6} mirrors the theory that led to the formulation of $\mathrm{P}_{\rm VI}$ by Fuchs via the classical theory of monodromy preserving deformations \cite{Fuchs1, Fuchs2}. The associated linear problem originally derived by Jimbo and Sakai~\cite{Sakai:qP6} is given by
\[
A(x) = \begin{pmatrix} \kappa_1((x-y)(x-\alpha) + z_1) & \kappa_2 w(x-y)\vspace{1mm} \\
\dfrac{\kappa_1}{w}(\gamma x + \delta) & \kappa_2((x-y)(x-\beta) +z_2)
\end{pmatrix},
\]
where letting the eigenvalues of $A(0)$ be $\lambda_1$ and $\lambda_2$, and the determinant of $A(x)$ be
\[
\det A(x) = \kappa_1\kappa_2 (x-a_1)(x-a_2)(x-a_3)(x-a_4),
\]
leads to
\begin{gather*}
\alpha = \dfrac{-z_1 \kappa _1-\left(y \left(-2 y+a_1+a_2+a_3+a_4\right)+z_2\right) \kappa _2+\lambda _1+\lambda _2}{y \left(\kappa _1-\kappa _2\right)},\\
\beta  = \dfrac{\left(y \left(-2 y+a_1+a_2+a_3+a_4\right)+z_1\right) \kappa _1+z_2 \kappa _2-\lambda _1-\lambda _2}{y \left(\kappa _1-\kappa _2\right)},\\
\gamma  =y^2+2 (\alpha +\beta ) y+\alpha  \beta -a_2 a_3-\left(a_2+a_3\right) a_4-a_1 \left(a_2+a_3+a_4\right)+z_1+z_2,\\
\delta  = -\dfrac{\left(y \alpha +z_1\right) \left(y \beta +z_2\right)-a_1 a_2 a_3 a_4}{y}.
\end{gather*}
We need to choose how to def\/ine $z$, which we take to be specif\/ied by
\begin{gather*}
z_1  = \dfrac{\left(y-a_1\right) \left(y-a_2\right)}{z},\qquad
z_2  = z \left(y-a_3\right) \left(y-a_4\right).
\end{gather*}
The theory regarding the existence and convergence of $Y_0(x)$ and $Y_{\infty}(x)$ for this case, where the asymptotics of the leading terms in the expansion of $A(x)$ around $x= 0$ and $x =\infty$ are invertible, was determined very early by Birkhof\/f et al.~\cite{Birkhoff}. In light of this, we solve for the f\/irst terms of $Y_0$ and $Y_\infty$ so that we may take the expansion to be
\begin{gather*}
Y_0(x) =  \left(\left(\begin{array}{c}  \dfrac{w \left(\left(y \alpha +z_1\right) \kappa _1-\left(y \beta +z_2\right) \kappa _2+\lambda _1-\lambda _2\right)}{2 \kappa _1} \vspace{1mm} \\
 \delta \end{array}\right.\right.\\
\phantom{Y_0(x) =}{}
\left.\left.\quad
 \begin{array}{c}
  \dfrac{w \left(\left(y \alpha +z_1\right) \kappa _1-\left(y \beta +z_2\right) \kappa _2-\lambda _1+\lambda _2\right)}{2 \kappa
   _1} \vspace{1mm}\\
 \delta  \end{array}\right) + O(x) \right)
 \begin{pmatrix} e_{q,\lambda_1}(x) & 0 \vspace{1mm}\\ 0 & e_{q,\lambda_2}(x) \end{pmatrix}, \\
 Y_\infty(x) =  \left(I + \dfrac{1}{x}\begin{pmatrix} \dfrac{q (y+\alpha )}{q-1} & -\dfrac{q w}{q \kappa _1-\kappa _2} \vspace{1mm}\\
 \dfrac{q \gamma  \kappa _1}{w \left(\kappa _1-q \kappa _2\right)} & \dfrac{q (y+\beta )}{q-1} \end{pmatrix} + O\left(\dfrac{1}{x^2}\right) \right)\begin{pmatrix}\dfrac{e_{q,\kappa_1}(x)}{ \theta_q\left(x\right)^2} & 0 \vspace{1mm}\\ 0 & \dfrac{e_{q,\kappa_2}(x)}{\theta_q\left(x\right)^2} \end{pmatrix}.
\end{gather*}
It is this parameterization in which we have a number of symmetries being introduced. We have the group of permutations on the $a_i$ generated by the elements
\begin{gather*}
s_1  : \ \left\{ \begin{array}{c c c c} \kappa_1 & \kappa_2 & \lambda_1 & \lambda_2 \\ a_1 & a_2 & a_3 & a_4 \end{array}; w, y,z \right\} \to
				\left\{ \begin{array}{c c c c} \kappa_1 & \kappa_2 & \lambda_1 & \lambda_2 \\ a_2 & a_1 & a_3 & a_4 \end{array}; w, y,z \right\}, \\
s_2  : \ \left\{ \begin{array}{c c c c} \kappa_1 & \kappa_2 & \lambda_1 & \lambda_2 \\ a_1 & a_2 & a_3 & a_4 \end{array}; w, y,z \right\} \to
				\left\{\begin{array}{c c c c} \kappa_1 & \kappa_2 & \lambda_1 & \lambda_2 \\ a_1 & a_3 & a_2 & a_4 \end{array}; w, y,z\dfrac{y-a_3}{y-a_2} \right\}, \\
s_3  : \ \left\{ \begin{array}{c c c c} \kappa_1 & \kappa_2 & \lambda_1 & \lambda_2 \\ a_1 & a_2 & a_3 & a_4 \end{array}; w, y,z \right\} \to
				\left\{ \begin{array}{c c c c} \kappa_1 & \kappa_2 & \lambda_1 & \lambda_2 \\ a_1 & a_2 & a_4 & a_3 \end{array}; w, y,z \right\} ,
\end{gather*}
which are all elementary B\"acklund transformations, as is the elementary transformation swapping the eigenvalues/asymptotic behavior at $x=0$, given by
\[
s_4 : \ \left\{ \begin{array}{c c c c} \kappa_1 & \kappa_2 & \lambda_1 & \lambda_2 \\ a_1 & a_2 & a_3 & a_4 \end{array}; w, y,z \right\} \to
				\left\{ \begin{array}{c c c c} \kappa_1 & \kappa_2 & \lambda_2 & \lambda_1 \\ a_1 & a_2 & a_3 & a_4 \end{array}; w, y,z \right\}.
\]
To obtain the corresponding permutation of $\kappa_1$ and $\kappa_2$ we multiply $Y(x)$ on the left by
\[
R(x) = \begin{pmatrix} 0 & 1 \\ 1 & 0 \end{pmatrix},
\]
in which the resulting transformation is most cleanly represented in terms of the parameterization of the linear problem, given as
\begin{gather*}
s_5 : \ \left\{ \begin{array}{c c c c} \kappa_1 & \kappa_2 & \lambda_1 & \lambda_2 \\ a_1 & a_2 & a_3 & a_4 \end{array}; w, y,z \right\}\\
\phantom{s_5 : \ }{} \ \to
\left\{ \begin{array}{c c c c} \kappa_2 & \kappa_1 & \lambda_1 & \lambda_2 \\ a_1 & a_2 & a_3 & a_4 \end{array};
  \dfrac{\kappa_1\gamma}{w} , -\dfrac{\delta}{\gamma},z \dfrac{(\delta + \gamma a_1)(\delta + \gamma a_2)}{(y\gamma + \delta)(\beta \gamma+ \delta) + \gamma^2z_2} \right\}.
\end{gather*}
Since the operator, $ T_{\kappa_1,\kappa_2,\lambda_1,\lambda_2}$, which is induced by $R_{\kappa_1,\kappa_2,\lambda_1,\lambda_2}(x) = xI$ is still a simple shift, if we include $ T_{\kappa_1,\kappa_2,\lambda_1,\lambda_2}$ and the symmetries $s_1, \ldots, s_5$, we still need two non-trivial translations, $T_{a_1,\lambda_1}$ and $T_{\kappa_1,\lambda_1}$. These are specif\/ied by
\begin{gather*}
T_{a_1, \lambda_1}  : \
\tilde{w} = w-\dfrac{w y \left(y-a_1\right) a_1 \left(q \kappa _1-\kappa _2\right)}{a_1 \left((z-1) y^2+\left(a_1-z \left(a_3+a_4\right)\right) y+z a_3 a_4\right) \kappa _2+\left(y-a_1\right) \lambda
   _2},\\
\phantom{T_{a_1, \lambda_1}  : \ }{} \
\tilde{y} = \dfrac{a_2 \kappa _1 (w-\tilde{w}) \left(a_3 a_4 \kappa _2 \tilde{z}-q \lambda _1\right)}{\lambda _1 \left(q \kappa _1 (\tilde{w}+w (\tilde{z}-1))-\kappa _2 \tilde{w} \tilde{z}\right)},\\
\phantom{T_{a_1, \lambda_1}  : \ }{} \
\tilde{z} =\dfrac{\kappa _1 \left(a_2 \left(q \left(w \left(y-a_1\right)+\tilde{w} \left(y (z-1)+a_1\right)\right) \kappa _1-w y z \kappa _2\right)+q z \left(w-\tilde{w}\right) \lambda _1\right)}{a_2
   \kappa _1 \left(q w \left(y-a_1\right) \kappa _1+\left(\tilde{w} \left(y (z-1)+a_1\right)-w y z\right) \kappa _2\right)+z \left(q w \kappa _1-\tilde{w} \kappa _2\right) \lambda _1},
\\
T_{\kappa_1,\lambda_1}  : \
\tilde{w}= \dfrac{q^2 w \left(\psi  a_1+1\right) \left(\psi  a_2+1\right) \kappa _1 \left(q \kappa _1-z \kappa _2\right) (\tilde{z}-1)}{z \psi  \left(\kappa _2-q \kappa _1\right)^2 \tilde{z}},\qquad
\tilde{y} =\dfrac{a_3 a_4 \kappa _2 \tilde{z}-q \lambda_1}{q \psi  \lambda _1-q \psi  \lambda _1 \tilde{z}},\\
\phantom{T_{\kappa_1,\lambda_1}  : \ }{} \
\tilde{z}z = \dfrac{q^2 \left(\psi  a_1+1\right) \left(\psi  a_2+1\right) \kappa _1 \lambda _1}{\left(a_3 \kappa _2+q \psi  \lambda _1\right) \left(a_4 \kappa _2+q \psi  \lambda _1\right)},
   \end{gather*}
where
\[
\psi = \dfrac{y \left(z \kappa _2-q \kappa _1\right)}{q \left(a_1 a_2 \kappa _1-z \lambda _1\right)}.
\]
The remaining elementary Schlesinger transformations are obtained from the symmetries and the above shifts. This gives
\begin{gather*}
\mathbb{Z}^7  \cong \langle T_{\kappa_1,\kappa_2,\lambda_1,\lambda_2}, T_{\kappa_1,\lambda_1} , T_{\kappa_1,\lambda_2},  T_{a_1,\lambda_1},T_{a_2,\lambda_1},T_{a_3,\lambda_1},T_{a_4,\lambda_1}  \rangle = G_A.
\end{gather*}

We may identify the evolution of $q$-$\mathrm{P}_{\rm VI}$ with the connection preserving deformation
\[
q\textrm{-}\mathrm{P}_{\rm VI} : \ \left\{ \begin{array}{c c c c} a_1 & a_2 & a_3 & a_4 \\ \kappa_1 & \kappa_2 & \lambda_1 & \lambda_2 \end{array} : w, y, z\right\} \to
\left\{ \begin{array}{c c c c} qa_1 & qa_2 & a_3 & a_4 \\ \kappa_1 & \kappa_2 & q\lambda_2 & q\lambda_1 \end{array} : \tilde{w}, \tilde{y}, \tilde{z}\right\},\\
\]
or in terms of Schlesinger transformations
\begin{gather*}
q\textrm{-}\mathrm{P}_{\rm VI} = T_{a_1,\lambda_1, a_2, \lambda_2}: \
\tilde{w} = \dfrac{q w \kappa _1 (\tilde{z}-1)}{\kappa _2 \tilde{z}-q \kappa _1},\qquad
\tilde{y}y = -\dfrac{\left(a_3 a_4 \kappa _2 \tilde{z}-q \lambda _1\right) \left(\lambda _1 \tilde{z}-q a_1 a_2 \kappa _1\right)}{\lambda _1 (\tilde{z}-1) \left(q \kappa _1-\kappa _2 \tilde{z}\right)},\\
\phantom{q\textrm{-}\mathrm{P}_{\rm VI} = T_{a_1,\lambda_1, a_2, \lambda_2}: \ }{} \
\tilde{z}z = \dfrac{q \left(y-a_1\right) \left(y-a_2\right) \kappa _1}{\left(y-a_3\right) \left(y-a_4\right) \kappa _2}.
\end{gather*}
We make a the correspondence with \eqref{qPVI} by letting $y = f$ and $z = g/b_3$, and the relations between parameters are given by
\begin{gather*}
a_1 = b_5, \qquad a_2 = b_6, \qquad  a_3 = b_7, \qquad a_4 = b_8, \qquad
\kappa_1 = \dfrac{b_4\lambda_1}{b_2b_7b_8}, \qquad \kappa_1 = \dfrac{b_3\lambda_1}{b_2b_7b_8},
\end{gather*}
which completes the correspondences. We remark that in the representation of \eqref{qPVI} the transformation $b_5 \to q b_5$, $b_6 \to qb_6$, $b_7\to qb_7$, $b_8 \to q b_8$ and $f \to q f$ is a trivial transformation and corresponds to $T_{a_1,a_2,a_3,a_4, \lambda_1^2,\lambda_2^2}$, which may also be regarded as trivial.

\section{Conclusion and discussion}\label{section4}

The above list comprises of $2\times 2$ associated linear problems for $q$-Painlev\'e equations in which the associated linear problem is quadratic in~$x$. The way in which we have increased the dimension of the underlying lattice of connection preserving deformations has been to successively increase the number of non-zero roots of the determinant. If we were to consider a natural extension of the $q$-$\mathrm{P}\big(A_3^{(1)}\big)$ case by adding a root making the determinant of order f\/ive in $x$ then we would necessarily obtain a matrix characteristically dif\/ferent from the Lax pair of Sakai~\cite{SakaiE6} and Yamada~\cite{Yamada:LaxqEs}. Sakai adjoins two roots of the determinant while introducing a relation between~$\kappa_1$ and~$\kappa_2$ while Yamada~\cite{Yamada:LaxqEs} introduces four roots on top of the $q$-$\mathrm{P}\big(A_3^{(1)}\big)$ case and f\/ixes the asymptotic behaviors at $x=0$ and $x=\infty$. This seems to skip the cases in which the determinant of $A$ is f\/ive. Of course, the above theory could be applied to those Lax pairs of Sakai and Yamada  \cite{SakaiE6, Yamada:LaxqEs}, yet, the jump between the given characteristic data for the associated linear problems of $q$-${\rm P}\big(A_3^{(1)}\big)$ and $q$-${\rm P}\big(A_2^{(1)}\big)$ is intriguing, and worth exploring in greater depth.

Asides from the exceptional cases, where the symmetry of the underlying discrete Painlev\'e equations are of type $E_6^{(1)}$, $E_7^{(1)}$ or $E_8^{(1)}$, there are a few issues that need to be addressed. We have outlined a very natural lattice structure underlying the connection preserving deformations, however, it is unknown whether the full set of symmetries may be derived from the underlying connection preserving deformation setting. It seems that in order to fully address this, one would require a certain duality between roots of the determinant and the leading behavior at~$0$ and~$\infty$. This could be explained from the point of view of rational matrices for example. We expect a framework which gives rise to a lift of the entire group of B\"acklund transformations to the level of Schlesinger transformations. This work, and the work of Yamada~\cite{Yamada:LaxqEs} may provide valuable insight into this.

We note the exceptional work of Noumi and Yamada in this case for a continuous analogue in their treatment of~$\mathrm{P}_{\rm VI}$ \cite{YamadaNoumi:LaxP6}. They are able to extend the work of Jimbo and Miwa~\cite{Jimbo:Monodromy2} for~$\mathrm{P}_{\rm VI}$ in a manner that includes all the symmetries of~$\mathrm{P}_{\rm VI}$. We remark however that the more geometric approach to Lax pairs explored by Yamada \cite{Yamada:E8Lax} gives us a~greater insight into a~possible fundamental link between the geometry and Lax integrability of the discrete Painlev\'e equations.

\pdfbookmark[1]{References}{ref}
\LastPageEnding

\end{document}